\documentclass[11pt]{article}

\def\be{\begin{equation}}
\def\ee{\end{equation}}
\def\bea{\begin{eqnarray}}
\def\eea{\end{eqnarray}}

\usepackage{hyperref}
\usepackage{upgreek}
\usepackage{bbm}
\usepackage{graphicx}
\usepackage{amsmath}
\usepackage{amssymb}

\usepackage{psfrag}
\usepackage{verbatim}
\usepackage[numbers, sort&compress]{natbib}
\usepackage[scale={.75,.75}]{geometry}
\usepackage{hyperref}

\newcommand{\slashed}[1]{\displaystyle{\not}{#1}}
\newcommand{\M}{{\rm M}}

\begin{document}
\date{}
\title{\vspace{-2.5cm} 
\begin{flushright}
\vspace{-0.4cm}
{\scriptsize \tt TUM-HEP-942/14}  
\end{flushright}
{\bf \footnotesize{HOW MANY NEW PARTICLES DO WE NEED AFTER THE HIGGS BOSON?}}}
\author{Marco Drewes \\
\footnotesize{Physik Department T70, Technische Universit\"at M\"unchen, }\\
\footnotesize{James Franck Stra\ss e 1, D-85748 Garching, Germany}
}
\maketitle
\date{}
\begin{abstract}
\noindent The discovery of the scalar boson completes the experimental confirmation
of the particles predicted by the Standard Model, which achieves to describe
almost all phenomena observed in nature in terms of a few symmetry
principles and a handful of numbers, the constants of nature. Neutrino
oscillations are the only confirmed piece of evidence for physics beyond
the Standard Model found in the laboratory. They can easily be explained
if the neutrinos have partners with right handed chirality like all
other fermions. Remarkably, right handed neutrinos can simultaneously
explain long standing puzzles from cosmology, namely Dark Matter and the
baryon asymmetry of the universe. I discuss how close this minimal
extension of the Standard Model by right handed neutrinos can bring us
to a complete theory of nature and what else may be needed.
\end{abstract}
\section{Introduction}
One way to define the ultimate goal of fundamental science is the much-anticipated \emph{theory of everything}. However, given the fact that we are limited in our ability to make observations by time, space and the sensitivity of our senses and instruments, it is clear from the beginning that this goal cannot be achieved - even if one day we can consistently describe all phenomena we know, we can never be sure that there are no phenomena which have escaped (and will always escape) our notice because they are too far away, involve very feeble interactions with the particles we know, occur only at very short distances or happened in a very distant past.
Hence, we may pursue a less ambitious, but nevertheless challenging and possibly achievable goal, a ``theory of everything we know'' or \emph{complete effective theory of nature}. Such a theory should 
\begin{itemize}
\item[1)] describe all phenomena observed in nature and
\item[2)] be testable experimentally.
\end{itemize}
We shall interpret condition 2) in a strong sense and demand that the existence of all particles can be confirmed experimentally and we can study their interactions. In the following we define a complete effective theory as one that fulfils both conditions.

It is immediately clear that there is no guarantee that such theory exists. If, for instance, neutrino masses are generated by physics at energy scales $\sim 10^{16}$ GeV, the (expected) scale of grand unification, then we will most likely never be able to study the mechanism behind this observed phenomenon experimentally. In fact, many popular and well-motivated theories do not fulfil criterion $2)$, including grand unification and most realisations of supersymmetry.
Hence, a theory as anticipated here can only be found if nature is kind enough that all phenomena within reach of our instruments can be explained by physics that is also within reach of our instruments. 
Even if this is the case the issue of a full (non-perturbative) theory of quantum gravity remains.\footnote{We have not observed any phenomenon that require such theory, hence we can formally pass both criteria without it.
On the other hand, a truly fundamental theory of nature (whatever that means) should include a quantum theory of gravity not only for self-consistency, but also because it seems likely that there exist physical systems for which 
it is required and which some day may be accessible to (at least indirect) observation, including black holes and the very early universe.} 
Leaving this aside, we in the following consider a theory as ``complete'' in the sense of 1) and 2)  and treat it as fundamental for all practical purposes if it can be a valid effective field theory up to the Planck scale.

The Standard Model of particle physics (SM) and theory of general relativity (GR) in combination come impressively close to fulfilling both conditions \cite{Beringer:1900zz}. 
To date, there are only four experimental and observational facts which 
cannot be understood in this framework,\footnote{The observed current acceleration of the universe's expansion is often included in this list, but can in fact be accommodated in the framework of SM+GR: All observations to date can be explained in terms of a cosmological constant $\Lambda$, which is simply a free parameter in GR. 
Whatever is the microphysical origin of $\Lambda$ lies outside our current observation, hence knowledge of it is not required to pass conditions $1)$ and $2)$.
}
\begin{itemize}
\item[(I)]  the spacetime geometry of the observable universe 
\item[(II)] flavour violation in neutrino experiments,\footnote{It is sometimes argued that flavour oscillations are ``not really'' physics beyond the SM, as they can be explained by simply adding a neutrino mass term. However, in a gauge invariant and renormalisable theory such a term can only arise if new physical degrees of freedom are added to the SM, see e.g. discussion in section 2 of Ref.~\cite{Drewes:2013gca}. The right handed neutrinos introduced in Sec.~\ref{CompleteTheory} are one way to achieve this, but not the only one (see e.g. \cite{Fukugita:2003en,Mohapatra:1998rq} for a detailed summary). } 
\item[(III)] the baryon asymmetry of the universe (BAU) and 
\item[(IV)] the composition and origin of the observed dark matter (DM).
\end{itemize}
In addition to this {\it evidence} for the existence of ``new physics'', there are a number of anomalies in experimental data that have not (yet?) led to a claim of discovery and may also be explained by systematics.
Moreover, there are aspects of the SM that can be considered unsatisfying from an aesthetic viewpoint, such as the hierarchy between the electroweak and Planck scale, the strong CP problem, the factorisation of the gauge group and the flavour structure. We do not discuss these here. Instead, we focus on one particular candidate for a complete theory, which is motivated by the principle of minimality or \emph{Ockham's razor}. 
\section{A ``complete'' theory of nature}\label{CompleteTheory}
All matter particles in the Standard Model 
(SM) except neutrinos have been observed with both, left-handed (LH) and right-handed (RH) chirality.
The lack of a RH counterpart implies that neutrinos are massless in the SM, while the observed neutrino flavour oscillations (II) clearly suggest that at least two neutrinos are massive.
This provides strong motivation to assume that RH neutrinos $\nu_R$ exist.
These new particles in addition can simultaneously explain the observed DM \cite{Dodelson:1993je,Shi:1998km} and BAU \cite{Fukugita:1986hr}.
It has been suggested in \cite{Asaka:2005pn} that they may indeed explain all observed phenomena (II)-(IV) simultaneously, for a review see e.g. Refs.~\cite{Boyarsky:2009ix,Abazajian:2012ys,Drewes:2013gca} and references therein, while the geometry (I) of the universe can be explained without adding any new particles \cite{Bezrukov:2007ep}.
We consider the most general action that only contains SM fields and 
RH neutrinos 
with renormalisable interactions
\begin{eqnarray}
	\label{L}
	\lefteqn{\mathcal{S} =\int d^4 \sqrt{-g} \bigg[\mathcal{L}_{SM}-\frac{\M^2}{2}R-\xi\Phi^\dagger\Phi R}\\ 
	&+&i \overline{\nu_{R}}\slashed{\partial}\nu_{R}-
	\overline{l_{L}}F\nu_{R}\tilde{\Phi} -
	\overline{\nu_{R}}F^{\dagger}l_L\tilde{\Phi}^{\dagger} 
        -{\rm \frac{1}{2}}(\overline{\nu_R^c}M_{M}\nu_{R} 
	+\overline{\nu_{R}}M_{M}^{\dagger}\nu^c_{R})\bigg]
        . \nonumber
	\end{eqnarray}
Here flavour and isospin indices are suppressed.
$\mathcal{L}_{SM}$ is the SM Lagrangian, $F$ is a matrix of
Yukawa couplings and $M_{M}$ a Majorana mass term for $\nu_{R}$. 
$l_{L}=(\nu_{L},e_{L})^{T}$ are the left handed SM lepton doublets, $\Phi$ is the Higgs doublet with 
$\tilde{\Phi}=(\epsilon\Phi)^\dagger$, where $\epsilon$ is the antisymmetric $SU(2)$ tensor, and $\nu_R^c=C\overline{\nu_R}^T$, with the charge conjugation matrix $C=i\gamma_2\gamma_0$.
For $n$ flavours of $\nu_R$, the eigenvalues of $M_M$ introduce $n$ new mass scales in nature, which we shall label $M_I$. In analogy with the LH sector we consider the case of $n=3$ flavours of RH neutrinos. 
This is the minimal number required to generate three non-zero light neutrino masses. 
We work in a flavour basis where $M_M={\rm diag}(M_1,M_2,M_3)$.
The action (\ref{L}) is written in the Jordan frame, $\M$ is a mass scale 
that can be related to (and almost equals) the Planck mass $M_P$ in the Einstein frame. 
The difference between Jordan and Einstein frame, which are connected by the conformal transformation $g_{\mu\nu}\rightarrow g_{\mu\nu}(\M^2+\xi\phi^2)/M_P^2$ with $\phi^2\equiv \Phi^\dagger\Phi$  \cite{Bezrukov:2008ut}, only matters in section \ref{sec:inflation}. Elsewhere we use the same symbols as in (\ref{L}) for the canonically normalised fields and follow the notation notation of Ref.~\cite{Drewes:2013gca}.
\section{The geometry of the universe and Higgs inflation}\label{sec:inflation}
If the universe contained only radiation and matter, then the cosmic microwave background (CMB) radiation from different directions in the sky would originate from regions that were causally disconnected at that time of emission. 
This makes it difficult to understand why the CMB temperature is the same up to Gaussian fluctuations $\delta T/T\sim 10^{-5}$ \cite{Smoot:1992td} in all directions  (``horizon problem'').
Moreover, the inferred overall spatial curvature is zero or very small \cite{Ade:2013zuv}, which means that it was extremely close to zero at earlier times (``flatness problem''). 
Both problems can be understood if there was a phase of \emph{cosmic inflation}, i.e. accelerated expansion, in the universe's very early history \cite{Starobinsky:1980te,Guth:1980zm,Linde:1981mu}.
Inflation also predicts Gaussian temperature/density perturbations \cite{Mukhanov:1981xt} with nearly flat spectrum, characterised by a spectral index $n_s$ close to one, in good agreement with observation.
In the model (\ref{L}) inflation can be realised through the potential energy of the Higgs field \cite{Bezrukov:2007ep}, which leads to a negative equation of state if it dominates the universe.
The Higgs expectation value $\chi$ in the Einstein frame is related to $\phi$ in the Jordan frame via the conformal transformation.
To explain the flatness and homogeneity of the universe inflation must last $\gtrsim 50-60$ $e$-folds, implying that the effective potential $U(\chi)$ must be sufficiently flat that $\chi$ ``rolls slowly'' while moving towards the minimum. Moreover, $U(\chi)$ must not have any wiggles at values below the scale of inflation.
If radiative corrections to $U(\chi)$ are negligible, the model (\ref{L}) is consistent with CMB observations if $\xi\simeq 47000 \sqrt{\lambda}$ and predicts a spectral index $n_s\simeq 0.97$ and scalar-to-tensor ratio $r\simeq0.003$ for CMB temperature fluctuations \cite{Bezrukov:2008ut}. Here $\lambda$ is the Higgs self-coupling.
In this scenario, hot big bang initial conditions for the radiation dominated era are generated through particle production and subsequent thermalisation during oscillations of $\chi$ around its minimum \cite{Bezrukov:2008ut}.
The reheating temperature is $10^{13}-10^{14}$ GeV, though this value is subject to some theoretical uncertainty, as the reheating process involves a complicated interplay between perturbative and non-perturbative dissipation \cite{GarciaBellido:2008ab}, and a detailed analysis requires a consistent treatment of all medium effects in the primordial plasma \cite{Drewes:2013iaa}. 
Radiative corrections introduce a critical mass scale \cite{Bezrukov:2008ej,DeSimone:2008ei,Barvinsky:2008ia,Bezrukov:2009db,Barvinsky:2009fy} 
$m_{\rm crit} = 129.6 + 2.0\frac{y_t-0.9361}{0.0058}
-0.5\frac{\alpha_s - 0.1184}{0.0007}$ GeV, where $y_t=y_t(\mu_t)$ is the top Yukawa at $\mu_t=173.2$ GeV in the $\overline{\rm MS}$ scheme and $\alpha_s$ is the strong coupling at the $Z$-mass. 
Inflation happens if the Higgs mass $m_H$ is larger than $m_{\rm crit}$ \cite{Bezrukov:2012sa,Degrassi:2012ry,Buttazzo:2013uya}.
While the above conclusions remain almost unchanged for $m_H> m_{\rm crit}$, calculations become rather sensitive to quantum corrections in the vicinity of $m_H=m_{\rm crit}$ \cite{Bezrukov:2014bra}. 
This uncertainty can be parametrised by two numbers that enter the relation between the inflationary and low energy values of the running masses. 
With the top quark mass varied within two standard deviations from its best fit experimental value, there exist values of these unknown parameters for which (\ref{L}) with $\xi\sim 10$, $r\gtrsim \mathcal{O}[10^{-1}]$ and $m_H=125.6$ GeV leads to Higgs driven inflation \cite{Bezrukov:2014bra,Hamada:2014iga}.
\section{Neutrino masses and the seesaw mechanism}
For $M_I>$  eV there are two distinct sets of neutrino mass eigenstates.
We represent them by flavour vectors of Majorana spinors $\upnu$ and $N$. 
The elements of $\upnu=V_\nu^{\dagger}\nu_L-U_\nu^{\dagger}\theta\nu_{R}^c +{\rm c.c}$ have light masses $\sim m_\nu=-\theta M_M \theta^T\ll M_M$ (in terms of eigenvalues) and are mainly superpositions of the ``active'' SU(2) doublet states $\nu_L$. 
They can be identified with the observed neutrino mass eigenstates.
The elements of $N=V_N^\dagger\nu_R+\Theta^{T}\nu_{L}^{c} +{\rm c.c}$ have masses of the order of $M_I$ and
are mainly superpositions of the ``sterile'' singlet states $\nu_R$. 
Here c.c. stands for the $C$-conjugation defined after (\ref{L}), 
$\Theta\ll \mathbbm{1}$ is the mixing matrix between active and sterile neutrinos and $\theta\equiv\Theta U_N^T$. 
$V_\nu$ is the usual neutrino mixing matrix and $U_\nu$ its unitary part, $V_N$ and $U_N$ are their equivalents in the sterile sector.
More precisely, $V_\nu\equiv (\mathbbm{1}-\frac{1}{2}\theta\theta^{\dagger})U_\nu$
with  $\theta\equiv m_D M_M^{-1}$,  $m_D\equiv F\chi$ ($\chi=174$ GeV at temperature $T=0$).
The unitary matrices $U_\nu$ and $U_N$ diagonalise the mass matrices $m_\nu\simeq-\theta M_M \theta^T$ and $M_N=M_M + \frac{1}{2}\big(\theta^{\dagger} \theta M_M + M_M^T \theta^T \theta^{*}\big)$, respectively. This setup is known as seesaw mechanism \cite{Minkowski:1977sc,GellMann:seesaw,Mohapatra:1979ia,Yanagida:1980xy}.

Experimentally the scale of the $M_I$ is almost unconstrained. Neutrino oscillation experiments at energies $E\ll M_I$ only involve the light states $\upnu_i$ and probe the specific combination $m_\nu=- FM_M^{-1}F^T$ \cite{Broncano:2002rw}.
If $m_H/M_I\ll 1$, then the $N_I$ tend to give large radiative corrections to $m_H$, and the observed light Higgs mass $m_H\sim 10^{-17}M_P$ \cite{Aad:2012tfa,Chatrchyan:2012ufa} 
can only be explained if either the model parameters are ``fine tuned'' or one introduces additional new physics to stabilise $m_H$ (e.g. supersymmetry).   
This \emph{hierarchy problem} is absent if $M_I<m_H$. We restrict ourselves to this regime here. 
Another motivation for this choice comes from the criterion $2)$: It is very difficult to find the $N_I$-particles in the laboratory in foreseeable time if they are heavier than the electroweak scale.

\section{Baryogenesis via leptogenesis}
\begin{figure}[h!]
\centering
\psfrag{mass}{$M_2$ [GeV]}
\psfrag{mixing}{$U_\mu^2$}
\includegraphics[width=9.5cm]{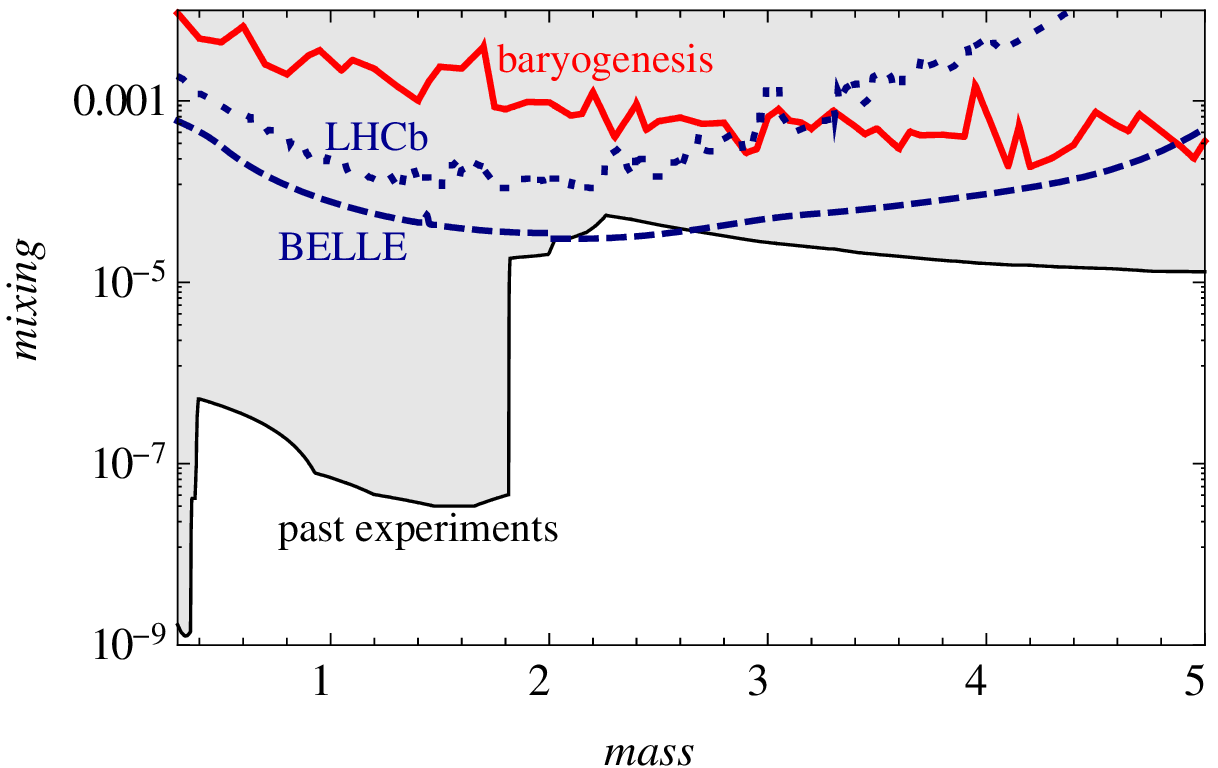}\\
\includegraphics[width=9.5cm]{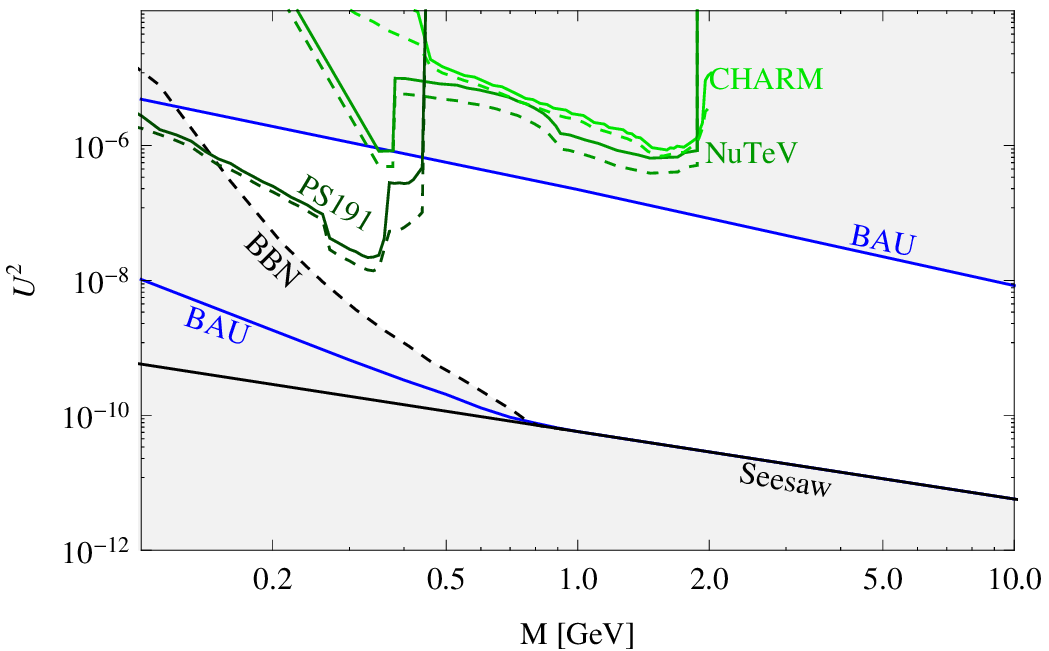}
\caption{Masses and mixings of $N_2$ for normal neutrino mass hierarchy.
\textbf{Upper panel}: For values of $|\Theta_{\mu 2}|^2$ below the red line the BAU can be explained for $n=3$ with $M_1=1$ GeV, $M_3=3$ GeV, $m_1 = 2.5\times 10^{-3}$ eV, 
$m_2 = 9.05 \times 10^{-3}$ eV $m_3 = 5 \times 10^{-2}$ eV and all known neutrino parameters fixed to the best fit value in Ref.~\cite{Capozzi:2013csa}. The gray area represents bounds on $U_\mu^2\equiv\sum_I|\Theta_{\mu I}|^2$ from the past experiments PS191 \cite{Bernardi:1987ek}, NuTeV \cite{Vaitaitis:1999wq} (both re-analysed in \cite{Ruchayskiy:2011aa}), NA3 \cite{Badier:1985wg}, CHARMII \cite{Vilain:1994vg} and DELPHI \cite{Abreu:1996pa} (as given in \cite{Atre:2009rg}). 
They are stronger than those from neutrinoless double $\beta$-decay \cite{Bezrukov:2005mx,LopezPavon:2012zg,Asaka:2013jfa,Merle:2013ibc,Girardi:2013zra,Ibarra:2011xn} or violation of lepton flavour \cite{Ibarra:2011xn,Canetti:2013qna} and universality \cite{Abada:2012mc,Abada:2013aba,Basso:2013jka,Endo:2014hza}.
The blue lines indicate the current bounds on $U_\mu^2$ from LHCb \cite{Aaij:2014aba} (dotted) and BELLE \cite{Liventsev:2013zz} (dashed). These can improve significantly (at least an order of magnitude) with the upgrade to BELLE II and the LHC's 14 TeV run, see discussion in Ref.~\cite{Canetti:2014dka}. For larger masses baryogenesis is possible \cite{Garbrecht:2014bfa}, but it is hard to find the $N_I$ in experiments.
\textbf{Lower panel}: With $n=2$ RH neutrinos baryogenesis can only be successful if their masses $M_2$ and $M_3$ are degenerate. 
The observed BAU can be generated for mixings $U^2\equiv{\rm tr}(\theta^\dagger\theta)$ in the region between the solid blue ``BAU'' lines. 
The regions below the solid black ``seesaw'' line and dashed black ``BBN''
line are excluded by neutrino oscillation experiments and big bang nucleosynthesis,
respectively. The areas above the green lines of different shade are
excluded by direct search experiments, as indicated in the plot. 
Plot taken from \cite{Canetti:2012kh}.
\label{LeptogenesisFig}
}
\end{figure}

The observable universe at present contains no significant amounts of antimatter, see \cite{Canetti:2012zc} for a discussion. In the standard model of cosmology the absence of antimatter is explained as the result of an almost complete mutual annihilation of matter and antimatter in the early universe after pair creation processes came out of thermal equilibrium. The matter we observe today is only a small remnant that survived this process due to a tiny excess $Y_B\simeq 8.6 \times 10^{-11}$ \cite{Ade:2013zuv} of matter over antimatter. 
The inflationary period described in section \ref{sec:inflation} and subsequent reheating produce a dense primordial plasma that contains matter and antimatter in equal amounts \cite{Bezrukov:2008ut}, hence $Y_B$ has to be generated dynamically at later times. This process of \emph{baryogenesis} requires baryon number ($B$)  violation, C and CP violation and a deviation from thermal equilibrium  \cite{Sakharov:1967dj}.
In the model (\ref{L}) the latter is realised during the production \cite{Akhmedov:1998qx,Asaka:2005pn} or the freezeout and decay \cite{Fukugita:1986hr} of $N_I$ in the early universe. 
We focus on the experimentally accessible mass range $M_I\sim$ GeV, in which the BAU is generated during $N_I$ production. This scenario is often referred to as \emph{baryogenesis from neutrino oscillations}. 
The asymmetry has to be generated at temperatures
$T> T_{\rm sph}\sim 130-140$ GeV \cite{Burnier:2005hp,D'Onofrio:2012jk,D'Onofrio:2014kta}, where sphaleron processes rapidly violate baryon number $B$ \cite{Kuzmin:1985mm}. 
The violation of total lepton $L$ number is suppressed by $M_I/T\ll 1$, but there can be significant asymmetries $Y_\alpha$ in the individual flavours. 
For $M_I/T \ll 1$ the helicity states of the Majorana fields $N_I$ effectively act as "particles" and "antiparticles", and one can assign approximately conserved lepton charges to the sterile flavours. 
Flavour dependent scatterings transfer a part $\delta L$ of the lepton asymmetry into the RH fields, where they are hidden from the sphaleron processes that partly transfer the remaining net asymmetry $-\delta L$ into $B$. 
Once the $N_I$ come into equilibrium the $Y_\alpha$ and $B$ get washed out. If this process is incomplete at the time of sphaleron freezeout at $T=T_{\rm sph}$, then a net $B\neq 0$ remains protected from further washout at lower temperatures. This mechanism is explained in more detail in Refs.~\cite{Boyarsky:2009ix,Canetti:2012kh,Drewes:2012ma,Drewes:2013gca,Khoze:2013oga,Shuve:2014zua}. Due to the great importance of flavour and finite density effects it is most conveniently treated in the nonequilibrium quantum field theory approach to leptogenesis \cite{Buchmuller:2000nd,DeSimone:2007rw,Garny:2009rv,Garny:2009qn,Anisimov:2010aq,Garny:2010nj,Beneke:2010wd,Garny:2010nz,Garbrecht:2010sz,Beneke:2010dz,Anisimov:2010dk,Fidler:2011yq,Garbrecht:2012qv,Garbrecht:2011aw,Garny:2011hg,Drewes:2012ma,Frossard:2012pc,Garbrecht:2014bfa}, though crucial results have previously been found in a detailed analysis using density matrix equations \cite{Asaka:2005pn}. The upper panel in figure \ref{LeptogenesisFig} shows that the 
masses and mixings required to explain the BAU with $n=3$ lie well within reach of the BELLE II and LHCb experiments. Hence, these experiments have the potential to unveil the common origin of matter in the universe and neutrino masses \cite{Canetti:2014dka}.

However, if one aims to address all problems (I)-(IV) in the framework of (\ref{L}), then one mass eigenstate $N_1$ must compose the observed DM and be very long lived. This implies that its coupling is so feeble that it gives no significant contribution to the neutrino mass matrix\footnote{This fixes the absolute scale of neutrino masses because it implies that one light neutrino is (almost) massless.} $m_\nu$  and has a negligible abundance in the primordial plasma at $T>T_{\rm sph}$. The latter fact implies that effectively only two RH neutrinos $N_{2,3}$ participate in baryogenesis. In this scenario, which is effectively $n=2$ as far as baryogenesis and the seesaw are concerned, it is much more difficult to explain the observed BAU.
The sources and washout rates for the asymmetries $Y_\alpha$ are proportional by different combinations 
of the same Yukawa couplings $F_{\alpha I}$.\footnote{The source term is e.g. given in \cite{Drewes:2012ma} and $\propto{\rm Im}(F_{\alpha I}F^*_{\beta I}F_{\beta J} F^*_{\alpha J})$, the washout rate is $\Gamma_\alpha=(F F^\dagger)_{\alpha\alpha}\gamma_{av}T$, where $\gamma_{av}$ is a numerical coefficient that depends on $M_I/T$ and has been calculated in thermal field theory in the relativistic and non-relativistic regime \cite{Anisimov:2010gy,Kiessig:2010pr,Garbrecht:2010sz,Laine:2011pq,Fidler:2011yq,Garbrecht:2011aw,Salvio:2011sf,Biondini:2013xua,Besak:2012qm,Garbrecht:2013bia,Bodeker:2014hqa,Laine:2013lka,Garbrecht:2013gd}.}
For $n=2$ the strengths of the $N_I$ couplings to all active flavours $\alpha$ are essentially governed by just one parameter \cite{Gorbunov:2007ak,Shaposhnikov:2008pf,Asaka:2011pb,Shuve:2014zua,Gorbunov:2013dta}. 
Hence, they are ``tied together'' and a large asymmetry generation at $T\gg T_{\rm sph}$ necessarily implies a large washout for all flavours at $T\gtrsim T_{\rm sph}$.\footnote{In contrast, for $n=3$ there are considerable regions in parameter space where the $|F_{\alpha I}|$ are very different in size, leading to a flavour asymmetric washout. For instance, the $N_I$ interactions with muons can be large enough to yield observable branching ratios (e.g. $({\rm tr}F^\dagger F)^{1/2}\gtrsim 10^{-4}$ at $M_2=2$ GeV) while the coupling to electrons is small enough to avoid a complete chemical equilibration \cite{Canetti:2014dka}.} 
The BAU can be explained if all $N_I$-interactions are sufficiently small to prevent a complete washout at $T>T_{\rm sph}$, see lower panel in figure \ref{LeptogenesisFig}, while the source term is resonantly enhanced by a degeneracy in the masses $M_2$ and $M_3$ at the level $<10^{-3}$ \cite{Canetti:2012kh,Shuve:2014zua}. A mass splitting of this size is stable against radiative corrections \cite{Roy:2010xq} and could be explained by an approximate lepton number conservation \cite{Shaposhnikov:2006nn}.
A detection of $N_{2,3}$ in existing experiments is unlikely in this scenario, and experimental confirmation of this model requires dedicated search experiments, see e.g. \cite{Bonivento:2013jag}. 
\section{Sterile neutrino Dark Matter} 
\begin{figure}[h!]
\begin{center}
\includegraphics[width=12cm]{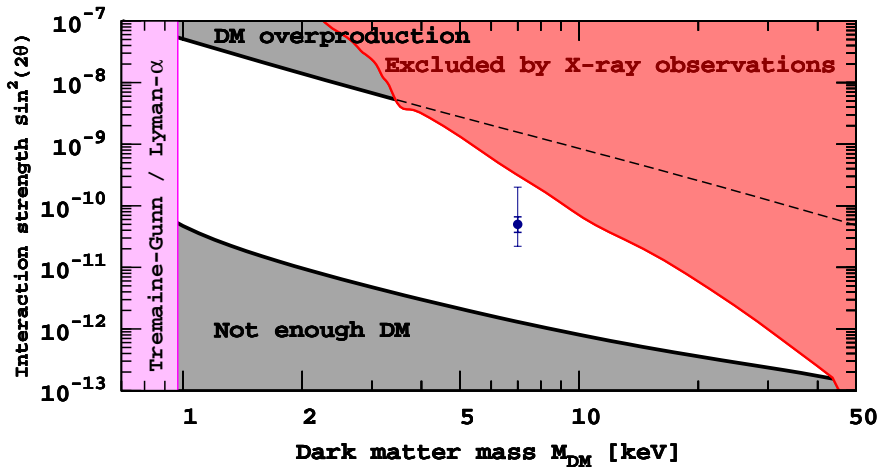}
\end{center}
\caption{Constraints on the mass and mixing of the DM candidate $N_1$ as explained in the plot.
Along the solid black \emph{production curves} the observed $\Omega_{DM}$ is explained for $Y_\alpha=0$ (upper curve) and the maximal $Y_\alpha=1.24\times 10^{-4}$ found in \cite{Canetti:2012kh} (lower curve).
We do not display a lower bound on $M_1$ from structure formation, which should lie somewhere between $1$ keV and $10$ keV, as it is a matter of ongoing discussion. Plot taken from \cite{Boyarsky:2014jta}.\label{DMfig}
}
\end{figure}

For sufficiently small $|\Theta_{\alpha I}|$, the $N_I$-particles are collisionless and can be very long lived, hence they are obvious decaying DM candidates. The main decay channel is $N\rightarrow \nu\nu\nu$ and leaves no astronomically observable signature, but the radiative decay $N\rightarrow\nu\gamma$ predicts a narrow photon emission line at energy $M_1/2$ from DM dense regions. Until 2014 the non-observation of such line could considerably constrain the mass and mixing \cite{Abazajian:2001vt,Boyarsky:2005us,Boyarsky:2006kc,Abazajian:2006yn,Abazajian:2006jc,Boyarsky:2006zi,Boyarsky:2006ag,Watson:2006qb,Riemer-Sorensen:2006pi,Boyarsky:2006fg,Boyarsky:2007ay,Boyarsky:2007ge,Yuksel:2007xh,Loewenstein:2008yi,RiemerSorensen:2009jp,Boyarsky:2009ix,Boyarsky:2010ci,Watson:2011dw,Essig:2013goa}, see figure \ref{DMfig}.
Recently a tentative signal has been reported that can be interpreted as emission from the decay of a sterile neutrino $N_1$ with $M_1\simeq 7$ keV \cite{Bulbul:2014sua,Boyarsky:2014jta}. This interpretation, which certainly requires further confirmation at this stage, 
is fully consistent with the predictions of the minimal model (\ref{L}) and all known constraints \cite{Canetti:2012vf,Boyarsky:2014jta,Abazajian:2014gza}.
The main constraints come from the requirements to predict the correct DM abundance $\Omega_{DM}$ and be consistent with structure formation, see e.g. \cite{Boyarsky:2009ix,Boyarsky:2012rt,Abazajian:2012ys,Drewes:2013gca} and references therein.
Even though $N_1$-particles in the universe are so feebly coupled that they have never reached thermal equilibrium, their distribution is proportional to an equilibrium distribution  if $Y_\alpha=0$ because they are thermally produced via mixing \cite{Dodelson:1993je,DiBari:1999ha,Dolgov:2000ew,Asaka:2006nq}. 
This production is most efficient at $T\sim 100$ MeV \cite{Asaka:2006nq,Asaka:2006rw}.  
For $Y_\alpha\neq0$ the MSW effect \cite{Wolfenstein:1977ue,Mikheev:1986gs} can lead to a level crossing between the $\upnu_i$ and $N_1$ dispersion relations in the thermal plasma, which results in a resonant producing of $N_1$ \cite{Shi:1998km} that adds a non-thermal component to the momentum distribution \cite{Laine:2008pg}. The superposition of the thermal and non-thermal spectra can be approximated as a combination of a warm and a cold DM component \cite{Laine:2008pg,Boyarsky:2008mt,Boyarsky:2008xj}. 
The observation of structures on scales $<100$ Mpc in the spatial matter distribution imposes a bound on the mean free path $\lambda_{DM}$ of $N_1$, as freely streaming DM would wash out and strongly suppress small scale structures. For a thermal spectrum there is a unique relation between $\lambda_{DM}$ and $M_1$, $\lambda_{DM}\sim 1 {\rm Mpc}({\rm keV}/M_1)$ \cite{Bond:1980ha}, but for a non-thermal spectrum it is very difficult to perform numerical simulations \cite{Lovell:2011rd} that make predictions for the distribution of matter in the universe today based on the initial $N_1$ spectra at $T\sim 100$ MeV, and the bounds from structure formation suffer from considerable uncertainties.
If one takes the most conservative viewpoint, then the $Y_\alpha=0$ scenario is already excluded (see e.g. \cite{Boyarsky:2009ix}) and the required $Y_\alpha\sim10^{-4}$ can only be produced in the late decay of $N_{2,3}$ in the minimal model discussed here if (\ref{L}) exhibits considerable parameter tunings \cite{Canetti:2012kh}.
Since the parameter space in figure \ref{DMfig} is constrained in all directions, the combination of observations with future X-ray telescopes \cite{Boyarsky:2012rt,Neronov:2013lqa,Takahashi:2012jn} and a better understanding of structure formation can in principle falsify or confirm this DM scenario.
\section{Conclusions}
In a minimal extension of the SM there exist parameter choices for which all established observations in fundamental physics can be explained with only three new particles, RH neutrinos, and a non-minimal coupling of the Higgs field to gravity. This model in principle could be a valid and complete effective field theory up to the Planck scale.
\bibliographystyle{JHEP}
\bibliography{all}
\end{document}